# Benchmarking OODBs with a Generic Tool

Jérôme Darmont, Michel Schneider

Laboratoire d'Informatique (LIMOS)
Université Blaise Pascal – Clermont-Ferrand II
Complexe Scientifique des Cézeaux
63177 Aubière Cedex
FRANCE

E-mail: *darmont@libd2.univ-bpclermont.fr, schneider@cicsun.univ-bpclermont.fr*

Phone: (33) 473-407-740
Fax: (33) 473-407-444

*Abstract*: *We present in this paper a generic object-oriented benchmark (OCB: the Object Clustering Benchmark) that has been designed to evaluate the performances of Object-Oriented Databases (OODBs), and more specifically the performances of clustering policies within OODBs. OCB is generic because its sample database may be customized to fit any of the databases introduced by the main existing benchmarks, e.g., OO1 (Object Operation 1) or OO7. The first version of OCB was purposely clustering-oriented due to a clustering-oriented workload, but OCB has been thoroughly extended to be able to suit other purposes. Eventually, OCB's code is compact and easily portable. OCB has been validated through two implementations: one within the $O_2$ OODB and another one within the Texas persistent object store. The performances of a specific clustering policy called DSTC (Dynamic, Statistical, Tunable Clustering) have also been evaluated with OCB.*

INTRODUCTION

The need to evaluate the performances of Object-Oriented Database Management Systems (OODBMSs) is important both to designers and users. Performance evaluation is useful to designers to determine elements of architecture, choose between caching strategies, and select Object Identifier (OID) type, among others. It helps them validate or refute hypotheses regarding the actual behavior of an OODBMS. Thus, performance evaluation is an essential component in the development process of efficient and well-designed object stores. Users may also employ performance evaluation, either to compare the efficiency of different technologies before selecting an OODBMS or to tune a system.

The work presented in this paper was initially motivated by the evaluation of object clustering techniques. The benefits induced by such techniques on global performances are widely acknowledged and numerous clustering strategies have been proposed. As far as we know, there is no generic approach allowing for their comparison. This problem is interesting for both designers (to set up the corresponding functionalities in the system kernel) and users (for performance tuning).

There are different approaches used to evaluate the performances of a given system: experimentation, simulation, and mathematical analysis. This paper focuses only on the first two approaches. Mathematical analysis is not considered because it invariably uses strong simplification hypotheses (Benzaken, 1990; Gardarin et al., 1995) and its results may well differ from reality.

Experimentation on the real system is the most natural approach and *a priori* the simplest to complete. However, the studied system must have been acquired, installed, and have a real database implanted in it. This database must also be significant of future exploitation of the system. Total investment and exploitation costs may be quite high, which can be drawbacks when selecting a product.

Simulation is casually used in substitution or as a complement to experimentation. It does not necessitate installing nor acquiring the real system. It can even be performed on a system still in development (*a priori* evaluation). The execution of a simulation program is generally much faster than experimentation. Investment and exploitation costs are very low. However, this approach necessitates the design of a functioning model for the studied system. The reliability of results directly depends on the quality and the validity of this model. Thus, the main difficulty is to elaborate and validate the model. A modelling methodology can help and secure these tasks.

Experimentation and simulation both necessitate a workload model (database and operations to run on this database) and a set of performance metrics. These elements are traditionally provided by benchmarks. Though interest for benchmarks is well recognized for experimentation, simulation approaches usually use workloads that are dedicated to a given study, rather than workloads suited to performance comparisons. We believe that benchmarking techniques can also be useful in simulation. Benchmarking can help validate a simulation model as compared to experimental results or support a mixed approach in which some performance criteria necessitating precision are measured by experimentation and other criteria that does not necessitate precision are evaluated by simulation.

There is no standard benchmark for OODBs, even if the more popular benchmarks, OO1, HyperModel, and OO7 are *de facto* standards. These benchmarks are aimed at engineering applications (Computer Aided Design, Manufacturing, or Software Engineering). These general-purpose benchmarks feature quite simple databases and are not well suited to the study of clustering, which is very data-dependent. Many benchmarks have been developed to study particular domains. A fair number of them are more or less based on OO1, HyperModel, or OO7.

In order to evaluate the performances of clustering algorithms within OODBs, we designed our own benchmark: OCB (Darmont et al., 1998). It originally had a generic object base and was clustering-oriented through its workload. It actually appeared afterwards that OCB could become more general, provided the focused workload was extended, as described in this paper.

The objective of this paper is to present full specifications for a new version of OCB. More precisely, we address the following points: the generalization of the OCB workload so that the benchmark becomes fully generic, a comparison of OCB to the main existing benchmarks, and a full set of experiments performed to definitely validate OCB. These performance tests were performed on the $O_2$ OODB (Deux, 1991), the Texas persistent object store (Singhal et al., 1992), and the DSTC clustering technique (Bullat & Schneider, 1996). The results obtained are discussed in this paper.

We are aware of the legal difficulties pointed out by Carey et al. (1993) and Carey et al. (1994). Indeed, OODBMS vendors are sometimes reluctant to see benchmark results published. The objective of our effort is rather to help software designers or users evaluate the adequacy of their product in a particular environment. OCB should also prove useful at least for researchers, to benchmark OODB prototypes and/or evaluate implementation techniques.

The remainder of this paper is organized as follows. The *de facto* standards in object-oriented benchmarking are briefly presented (OO1, HyperModel, and OO7; as well as the Justitia benchmark, which is interesting due to its multi-user approach). Next, our proposed benchmark, OCB, is described and compared to the other benchmarks. Experiments performed to validate our benchmark are also presented. We conclude the paper with future research directions.

RELATED WORK

The OO1 Benchmark

OO1, also referred to as the "Cattell Benchmark" (Cattell, 1991), was developed early in the 1990's when there was no appropriate benchmark for engineering applications. OO1 is a simple benchmark that is very easy to implement. It was used to test a broad range of systems including object-oriented DBMS, relational DBMS, and other systems such as Sun's INDEX (B-tree based) system. The visibility and simplicity of OO1 provide a standard for OODB benchmarking. A major drawback of this tool is that its data model is too elementary to measure the elaborate traversals that are common in many types of object-oriented applications, including engineering applications. Furthermore, OO1 only supports simple navigational and update tasks and has a limited notion of complex objects (only one composite hierarchy).

The HyperModel Benchmark

The HyperModel Benchmark (Anderson et al., 1990), also referred to as the Tektronix Benchmark, is recognized for the richness of the tests it proposes. HyperModel possesses both a richer schema and a wider extent of operations than OO1. This renders it potentially more effective than OO1 in measuring the performances of engineering databases. However, this added complexity also makes HyperModel harder to implement, especially since its specifications are not as complete as OO1's. The presence of complex objects in the HyperModel Benchmark is limited to a composition hierarchy and two inheritance links. The scalability of HyperModel is also not clearly expressed in the literature, whereas other benchmarks explicitly support different and well identified database sizes.

The OO7 Benchmark

OO7 (Carey et al., 1993) is a more recent benchmark than OO1 and HyperModel. It reuses their structures to propose a more complete benchmark and to simulate various transactions running on a diversified database. It has also been designed to be more generic than its predecessors and to correct their weaknesses in terms of object complexity and associative accesses. This is achieved with a rich schema and a comprehensive set of operations.

However, if OO7 is a good benchmark for engineering applications, it is not the case for other types of applications such as financial, telecommunication, and multimedia applications (Tiwary et al., 1995). Since its schema is static, it cannot be adapted to other purposes. Eventually, the database structure and operations of OO7 are nontrivial. Hence, the benchmark is quite difficult to understand, adapt, or even implement. Yet, to be fair, OO7 implementations are available by anonymous FTP[1].

The Justitia Benchmark

Justitia (Schreiber, 1994) has been designed to address the shortcomings of existing benchmarks regarding multi-user functionality, which is important in evaluating client-server environments. Justitia is also aimed at testing OODB capacity in reorganizing its database. Because Justitia's published specifications lack precision, the author's work cannot be easily reused. Furthermore, taking multiple users into account renders the benchmark quite complex. Justitia is fairly tunable and supposed to be generic, but it still uses structures that are typical of engineering applications. Its database schema is more limited than those of HyperModel or OO7. Though the object types are diverse, inter-class relationships are very few. The inheritance graph is substantial, but other types of references are limited to composition.

THE OBJECT CLUSTERING BENCHMARK

Originally, the purpose of OCB was to test the performances of clustering algorithms within object-oriented systems. OCB is structured around a rich object base including many different classes and numerous types of references allowing the design of multiple interleaved hierarchies. This database is wholly generic. The OCB workload, once clustering-oriented, has been extended with relevant, significant, and reproducible transactions. Thus, the workload became fully generic.

The flexibility of OCB is achieved through an extensive set of parameters. Many different kinds of object bases can be modeled with OCB as well as many different kinds of applications running on these databases. This is an important feature since there exists no canonical OODB application. OCB can indeed be easily parameterized to model a generic application or dedicated to a given type of object base and/or application. OCB is also readily scalable in terms of size and complexity resulting in a wide range of object bases. Usage time can be set up as well to be rather short or more extensive. Moreover, OCB's parameters are easy to set up.

---

[1] ftp://ftp.cs.wisc.edu/OO7

OCB's code is very compact and easily implemented on any platform. OCB is currently implemented in C++ to benchmark $O_2$ and Texas. Both versions are freely available[2]. The C++ code is less than 1,500 lines long. OCB has also been ported into QNAP2 and C++ simulation models. QNAP2 is a simulation software that supports a non object-oriented language close to Pascal. The QNAP2 code dealing with OCB is shorter than 1,000 lines.

The next version of OCB, which is currently in development, shall support multiple users viewed as processes in a very simple way to test the efficiency of concurrency control. As far as we know, Justitia is the only benchmark to have actually addressed this problem, though OO7 also has a multi-user version in development. OO1 was designed as multi-user, but the published results only involve a single user. One of our research objectives is to provide clear specifications for our benchmark so that others can readily implement it and provide feedback to improve it.

OCB Database

The OCB database is highly generic because it is rich, simple to achieve, and very tunable. It is made of a predefined number of classes (*NC*) derived from the same metaclass (Figure 1). A class has a unique logical identifier, *Class_ID*, and is defined by two parameters: *MAXNREF*, the maximum number of references in the class' instances; and *BASESIZE*, an increment size used to compute the *InstanceSize* after the inheritance graph is processed at database generation time. On Figure 1, note that the UML « bind » clause indicates that classes are instantiated from the metaclass using the parameters between brackets.

Since different references can point to the same class, 0-N, 1-N, and M-N links are implicitly modeled. Each of these *CRef* references has a type: *TRef*. There are *NTREF* different types of references. A reference type can be, for instance, a type of inheritance, aggregation, or user association. Eventually, an *Iterator* is maintained within each class to save references toward all its instances.

Objects in the database (instances of class *OBJECT*) are characterized by a unique logical identifier *OID* and by their class through the *ClassPtr* pointer. Each object possesses *AT-*

*TRANGE* integer attributes that may be read and updated by transactions. A string of size *InstanceSize*, the *Filler*, simulates the actual size the object should occupy on disk.

After instantiating the database schema, an object *O* of class *C* points through the *ORef* references to at most *MAXNREF* objects. These objects are selected from the iterator of the class referenced by *C* through the corresponding *CRef* reference. For each direct reference identified by an *ORef* link from an object $o_i$ toward an object $o_j$, there is also a backward reference (*BackRef*) from $o_j$ to $o_i$.

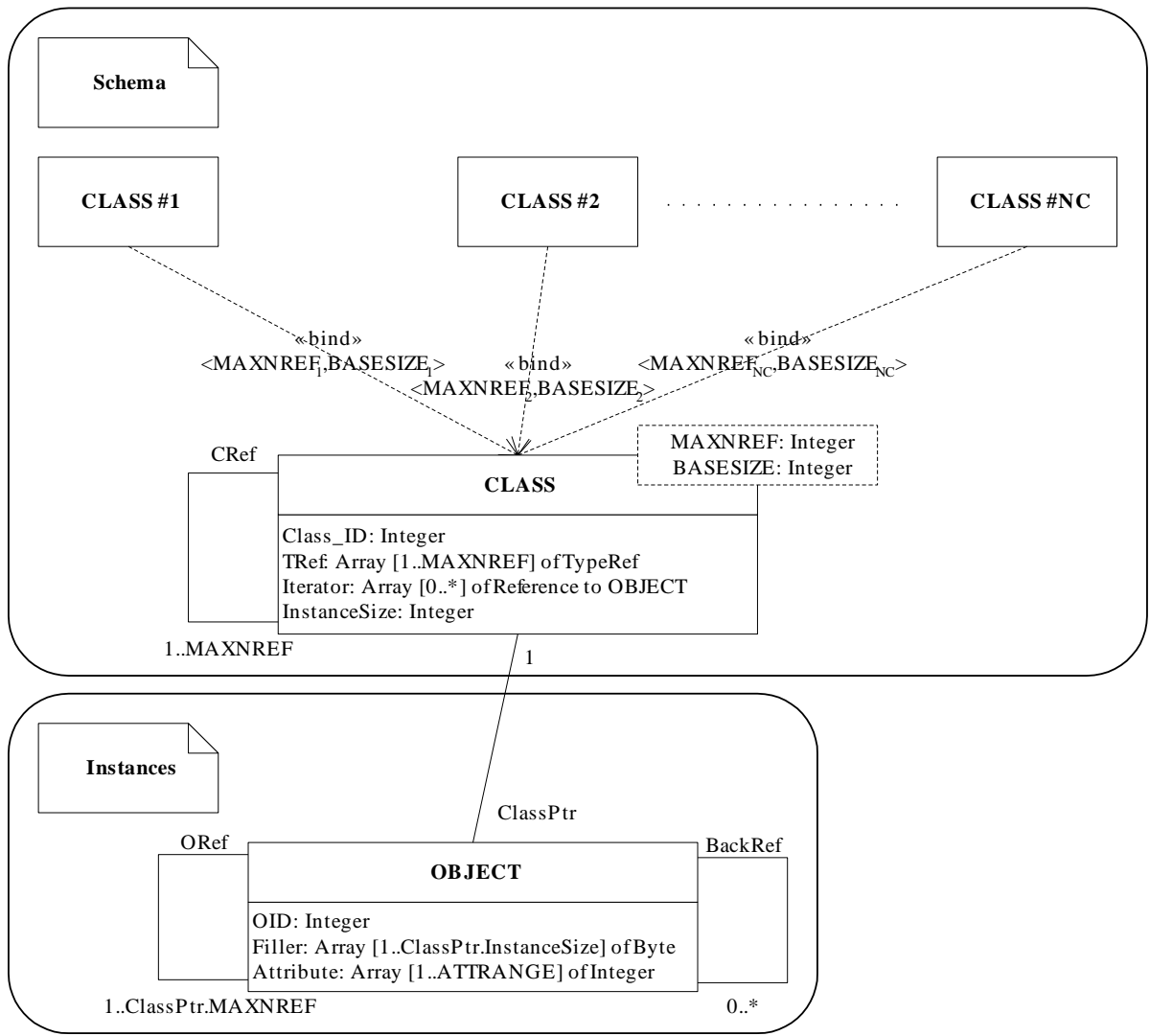

*Figure 1: OCB database schema (UML Static Structure Diagram)*

The database generation proceeds through three primary steps.

---

[2] http://libd2.univ-bpclermont.fr/~darmont/download/

1) Instantiation of the *CLASS* metaclass into *NC* classes: creation of the classes without any reference, then selection of the classes referenced by each class. The type of the references (*TRef*) can either follow the *DIST1* random distribution or be set up *a priori*. The referenced classes belong to the [*Class_ID – CLOCREF, Class_ID + CLOCREF*] interval that models a certain locality of reference, as introduced by OO1, but at the class level. The class reference selection can either follow the *DIST2* random distribution or be set up *a priori*. NIL references are possible.

2) Database consistency check-up: suppression of all the cycles and discrepancies within the graphs that do not allow them, e.g., inheritance graphs or composition hierarchies.

3) Instantiation of the *NC* classes into *NO* objects: creation of the objects, without any reference — their class follows the *DIST3* random distribution, then random selection of the objects referenced by each object. The referenced objects belong to the [*OID – OLOCREF, OID + OLOCREF*] interval that models a certain locality of reference at the instance level. The random selection of object references follows the *DIST4* random distribution. Reverse references (*BackRef*) are instantiated when the direct links are instantiated.

The random numbers are generated by the Lewis-Payne random generator (Lewis & Payne, 1973), which is one of the best pseudorandom number generators currently available. The database parameters are summarized in Table 1.

| Parameter name | Parameter | Default value |
|---|---|---|
| NC | Number of classes in the database | 50 |
| MAXNREF (i) | Maximum number of references, per class | 10 |
| BASESIZE (i) | Instances base size, per class | 50 bytes |
| NO | Total number of objects | 20,000 |
| NREFT | Number of reference types (inheritance, aggregation, etc.) | 4 |
| ATTRANGE | Number of integer attributes in an object | 1 |
| CLOCREF | Class locality of reference | NC |
| OLOCREF | Object locality of reference | NO |
| MAXRETRY | Maximum number of retries when linking objects | 3 |
| DIST1 | Reference types random distribution | Uniform |
| DIST2 | Class references random distribution | Uniform |
| DIST3 | Objects in classes random distribution | Uniform |
| DIST4 | Objects references random distribution | Uniform |

*Table 1: OCB database parameters*

OCB Workload

The core of the workload is organized around several transactions, the *traversals*, which are able to explore the effects of clustering. Several operations that do not benefit from any clustering effort have been re-introduced, e.g., creation and update operations. A full description of the benchmark's operations follows.

- *Random Access:* Access to *NRND* objects following the *DIST5* random distribution.
- *Sequential Scan:* Browse the instances of a class following the *DIST6* random distribution (*Simple Scan*). A *Range Lookup* additionally tests the value of *NTEST* attributes in each instance.
- *Traversals:* Traversal operations are divided into two types: *Set-Oriented Accesses* (or *Associative Accesses*) and *Navigational Accesses*, which have been empirically found by Mc Iver (1994) to match breadth-first and depth-first traversals; respectively. *Navigational Accesses* are further divided into *Simple*, depth-first traversals, *Hierarchy Traversals* that always follow the same type of reference, and finally *Stochastic Traversals* that randomly select the next link to cross. Stochastic traversals effectively simulate the access patterns caused by real queries, according to Tsangaris & Naughton (1992). An object bears at most *MAXNREF* references numbered from 1 to *MAXNREF*. At each step of a stochastic traversal, the probability to follow reference number $N$ ($N \in [1, MAXNREF]$) is $p(N) = \frac{1}{2^N}$. Each type of traversal proceeds from a root object following the *DIST7* random distribution and up to a predefined depth depending on the traversal type. All these transactions can be reversed to follow the links backward, "ascending" the graphs.
- *Update:* Update operations are also subdivided into different types. *Schema Evolutions* deal with individual insertion and deletion of *Class* objects. The class to be deleted follows the *DIST8* random distribution. *Database Evolutions* deal with individual insertion and deletion of objects. The object to be deleted follows the *DIST9* random distribution. Eventually, *Attribute Updates* allow attribute changes, either of random accessed objects (*Random Update* of *NUPDT* objects following the *DISTA* random distribution) or of instances of a class following the *DISTB* random distribution (*Sequential Update*).

The execution of transactions by each client (the benchmark is to be multi-user) is organized according to the following protocol:

1) cold run of *COLDN* transactions whose types are determined randomly according to pre-defined probabilities. The purpose of this step is to fill in the cache in order to observe the real, stationary behavior of the clustering algorithm implemented in the benchmarked system;

2) warm run of *HOTN* transactions.

A latency time *THINK* can be introduced between each transaction run. Furthermore, the whole benchmark execution may be replicated so that the same set of transactions is executed on different randomly-generated object bases. This feature allows the computation of mean values and confidence intervals, which are typically more significant than a single measurement. The OCB workload parameters are summarized in Table 2.

| Parameter(s) name(s) | Parameter(s) | Default value(s) |
|---|---|---|
| NRND | Number of objects accessed in Random Accesses | 50 |
| NTEST | Number of attributes tested in Range Lookups | 1 |
| SETDEPTH, SIMDEPTH, HIEDEPTH, STODEPTH | *Depth:* Set-oriented Access, Simple Traversal, Hierarchy Traversal, Stochastic Traversal | 3, 3, 5, 50 |
| NUPDT | Number of updated objects in Random Updates | 50 |
| DIST5, DIST6, DIST7, DIST8, DIST9, DISTA, DISTB | *Random distribution law:* Random Access objects, Sequential Scan classes, Transaction root objects, Schema Evolution classes, Database Evolution objects, Random Update objects, Sequential Update classes | Uniform |
| PRND, PSCAN, PRANGE, PSET, PSIMPLE, PHIER, PSTOCH, PCINSERT, PCDEL, POINSERT, PODEL, PRNDUP, PSEQUP | *Occurrence probability:* Random Access, Simple Scan, Range Lookup, Set Access, Simple Traversal, Hierarchy Traversal, Stochastic Traversal, Class Insertion, Class Deletion, Object Insertion, Object Deletion, Random Update, Sequential Update | 0.1, 0.05, 0.05, 0.2, 0.2, 0.2, 0.1, 0.005, 0.005, 0.02, 0.02, 0.025, 0.025 |
| COLDN | Number of transactions executed during the cold run | 1,000 |
| HOTN | Number of transactions executed during the warm run | 10,000 |
| THINK | Average latency time between two transactions | 0 |
| CLIENTN | Number of clients | 1 |
| RSEED | Random generator seed | Default seed |

*Table 2 : OCB workload parameters*

The metrics measured by OCB are basically:

- database response time (global and per transaction type) and throughput. In a client-server environment, times must be measured on the client side with standard system primitives like `time()` or `getrusage()` in C++. The replication of the transactions compensates for the possible inaccuracy of these functions. If the number of transactions is sufficiently large, the absence of such system functions can be compensated by a manual timing, as it is specified for OO1;

- number of accessed objects (both globally and per transaction type). The computation of these usage statistics must be included in the benchmark's code;
- number of Input/Output (I/Os) performed. The I/Os necessary to execute the transactions and the I/Os needed to cluster the database (clustering overhead) must be distinguished. I/O usage can be obtained through the C++ `getrusage()` function or by statistics provided by the DBMS. For instance, $O_2$ provides such statistics.

Comparison of OCB to the Existing Benchmarks

*Genericity of OCB*

Since we intend to provide a generic benchmark, our tool must be able to model various types of databases and applications. In addition, it must also be able to imitate the demeanor of previous object-oriented benchmarks. Schreiber (1994) claims Justitia bestows this property provided the benchmark is properly parameterized. However, he does not provide any solid evidence to back up his claim.

We have shown that the OCB database is generic by comparing it to the object bases from existing benchmarks (Tables 3 and 4). In terms of workload, however, the demonstration of genericity is more difficult to achieve. OO7 especially offers a wide range of complex transactions. Some of them have been discarded when designing OCB, because they added complexity without providing much insight. Still, the transactional behavior of OO1, HyperModel, and Justitia can easily be imitated. Furthermore, some of OCB's operations, if combined, can be equivalent to OO7's complex operations.

*Comparison with Gray's Criteria*

Gray (1993) defines four primary criteria concerning the specification of a good benchmark:
1) *relevance:* it must concern aspects of performance that appeal to the largest number of potential users;
2) *portability:* it must be reusable to test the performances of different OODBs;
3) *simplicity:* it must be feasible and must not require too many resources;
4) *scalability:* it must be able to be adapted to small or large computer systems, or new architectures.

| OCB parameter | OO1 | HyperModel |
|---|---|---|
| NC | 2 | 3 |
| MAXNREF (i) | Parts: 3<br>Connections: 2 | 5 (*Parent/Children*)<br>+ 5 (*PartOf/Part*)<br>+ *NO* (*RefTo/RefFrom*)<br>+ 1 (*Specialization*) |
| BASESIZE (i) | Parts: 50 bytes<br>Connections: 50 bytes | Node: 20 bytes<br>TextNode: 1000 bytes<br>FormNode: 20008 bytes |
| NO | 20000 parts<br>+ 60000 connections | 3906 Nodes<br>+ 125 FormNodes<br>+ 15500 TextNodes |
| NREFT | 3 | 4 |
| CREFLOC | *NC* | *NC* |
| OREFLOC | *RefZone* | Level *k+1* in the *Parent/Children* hierarchy |
| DIST1 | Constant (non random) | Constant (non random) |
| DIST2 | Constant (non random) | Constant (non random) |
| DIST3 | Constant (non random) | Constant (non random) |
| DIST4 | Uniform | Uniform |

*Table 3: OCB tuning to imitate OO1 and HyperModel object bases*

When designing OCB, we mainly intended to palliate two shortcomings in existing benchmarks: their lack of genericity and their inability to properly evaluate the performances of object clustering techniques. To achieve this goal, we designed a fully tunable benchmark, allowing it either to be generic or to be specialized for a given purpose. The consequences of this choice on Gray's criteria are the following:

- *relevance:* as previously stated, all the transactions from existing benchmarks have been included in OCB except the most intricate operations from OO7;
- *portability:* OCB has been used to evaluate the performances of the $O_2$ and the Texas systems. Both these implementations have been made in C++. OCB has also been included in simulation models written in QNAP2 and a simulation package called DESP-C++. OCB's code is short and simple in all these cases;
- *simplicity:* complete specifications for our benchmark are provided in this section in order to support understandability and ease of implementation;
- *scalability:* OCB is a very flexible benchmark due to an extensive set of parameters. Its object base can take different sizes and complexity levels and its various transactions can model a fair number of applications.

The characteristics of the existing benchmarks and OCB according to these criteria are summarized in Table 5.

*Benchmarking OODBs with a Generic Tool* – Submission to JDM                                    12/22

| OCB parameter | OO7 | Justitia |
|---|---|---|
| NC | 10 | 6 |
| MAXNREF (i) | Design object: 0<br>Atomic part: 20<br>Connection: 18<br>Composite part: *NumAtomicPerComp* + 8<br>Document: 1<br>Manual: 1<br>Assembly: 2<br>Complex assembly: *NumAssmPerAssm* + 2<br>Base assembly: *NumComPerAssm* x 2 + 1<br>Module: $\sum_{i=0}^{NumAssmLevels} NumAssmPerAssm^i$ | Database Entry: 0<br>Node: 2<br>CO: 3<br>PO: *PO_ATT_SIZE* + 3 |
| BASESIZE (i) | Design object: 18 bytes<br>Atomic part: 12 bytes<br>Connection: 14 bytes<br>Composite part: 0<br>Document: *DocumentSize* + 44 bytes<br>Manual: *ManualSize* + 48 bytes<br>Assembly: 0<br>Complex assembly: 0<br>Base assembly: 0<br>Module : 0 | Database entry: 4 bytes<br>PO: 0<br>Node: 4 bytes<br>CO: 0<br>DO: *DO_ATT_SIZE* bytes<br>SO: *SO_ATT_SIZE* bytes |
| NO | *NumModules* modules<br>+ *NumModules* manuals<br>+ $\sum_{i=0}^{NumAssmLevels-1} NumAssmPerAssm^i$ complex assemblies<br>+ *NumPerAssm*^*NumAssmLevels* base assemblies<br>+ *NumCompPerModule* composite parts<br>+ *NumCompPerModule* documents<br>+ *NumAtomicPerComp* . *NumCompPerModule* atomic parts<br>+ *NumAtomicPerComp* . *NumCompPerModule* . *NumConnPerAtomic* connections | *SECTION . MAXWIDTH . MAXLEVEL* |
| NREFT | 12 | 3 |
| CREFLOC | *NC* | *NC* |
| OREFLOC | *NO* | *NO* |
| DIST1 | Constant (non random) | Constant (non random) |
| DIST2 | Constant (non random) | Constant (non random) |
| DIST3 | Constant (non random) | Constant (non random) |
| DIST4 | Constant + Uniform | Constant (non random) |

*Table 4: OCB tuning to imitate OO7 and Justitia object bases*

|  | *Relevance* | *Portability* | *Simplicity* | *Scalability* |
|---|---|---|---|---|
| **OO1** | – – | ++ | ++ | – |
| **HyperModel** | + | + | – | – – |
| **OO7** | ++ | + | – | – |
| **Justitia** | – | – – | + | + |
| **OCB** | ++ | + | + | ++ |

*Strong point : +    Very strong point : ++    Weak point : –    Very weak point : – –*

*Table 5: Comparison of existing benchmarks to OCB*

VALIDATION EXPERIMENTS

We present in this section performance evaluations performed with OCB on the $O_2$ OODB, the Texas persistent object store, and the DSTC clustering technique, which is implemented in Texas. Our research objective did not include a comparison of the performances of $O_2$ and Texas. This would have been troublesome since our versions of these systems did not run on the same platform. Furthermore, $O_2$ and Texas are quite different in their philosophy and functionalities. $O_2$ is a full OODB supporting concurrent and secure accesses while Texas is positioned as an efficient persistent store for C++. We only intended to show that OCB provided valid performance evaluations.

Since we recommended the use of a complex object base, the feasibility of our specifications has been checked by measuring the database average generation time function of the database size (number of classes and number of instances). For schemas containing 10, 20, and 50 classes, the number of instances *NO* was varied from 5,000 to 50,000. The actual database size was also measured for all these configurations.

Next, the object base configuration was varied: number of classes *NC*, number of instances *NO*, number of inter-object references *MAXNREF*. Four database configurations were obtained using *NC* values of 20 and 50, and *MAXNREF* values of 5 and 10. Then, the number of instances in the database was varied from 500 to 20,000 for each configuration. The scope of this study is limited raw performance results, i.e., the average response time and the average number of I/Os necessary to execute the operations.

The efficiency of the DSTC clustering technique has been assessed by measuring the performances achieved by Texas before and after object clustering, on a medium and on a large database. The medium database was OCB's default object base: 50 classes, 20,000 instances, about 20 MB with Texas. Technical problems were encountered with Texas/DSTC to cluster a large database. The problem was circumvented by reducing the amount of available memory so that the database size was actually big compared to the size of the memory. To observe a significant gain in performances, DSTC was placed in advantageous conditions by running very characteristic transactions (hierarchy traversals and simple traversals from predefined root objects).

<u>Note</u>: All our experiments have been replicated 100 times so that mean tendencies could be assessed.

Results for $O_2$

*Material Conditions*

The $O_2$ server (version 5.0) was installed on an IBM RISC 6000 43P240 biprocessor workstation. Each processor was a Power PC 604e 166. The workstation had 1 GB ECC RAM. The operating system was AIX version 4. The $O_2$ server cache size was 16 MB by default.

*Object Base Generation*

Figure 2 displays the database generation time function of the number of classes and the number of instances in the base. It shows that generation time increased linearly when the schema was made of 10 or 20 classes. The increase was more accentuated with 50 classes, because when the $O_2$ client cache was full, which happened with the biggest databases, an execution error occurred. To fix this problem, the generation process has been marked out with commits. These multiple commits ware more costly than a single validation at the end of the generation process. The feasibility of OCB was also demonstrated, since in the worst case generation time was less than one hour. Moreover, a given object base could be saved and reused multiple times so that the generation process could be avoided each time.

Figure 3 shows how the size of the randomly generated database linearly evolved with the number of classes and instances. Hence, it was easy to foresee the final size of a database when setting the *NC* and *NO* parameters. The default OCB database (50 classes, 20,000 instances) had a mean size of 30 MB, which is average for a benchmark. For instance, the large database in OO1 has a size of 40 MB. However, we showed that larger databases are possible.

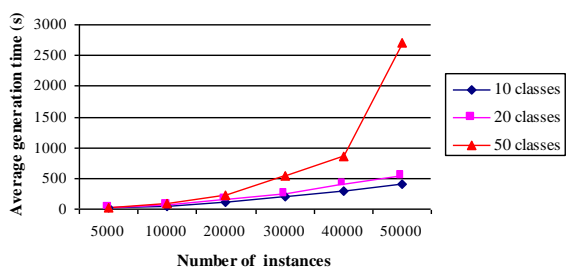
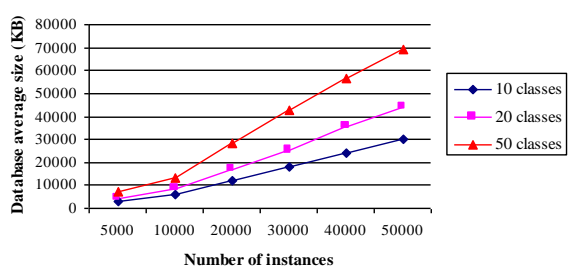

*Figure 2: Database generation time ($O_2$)*     *Figure 3: Actual database size ($O_2$)*

*Object Base Usage*

In Figure 4, we plotted the mean number of I/Os globally necessary to execute the transactions function of the number of instances in the object base (*NO*) for our four database configurations. We did the same in Figure 5 for the mean response time.

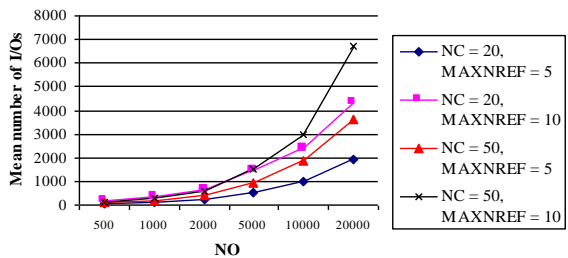
*Figure 4: Mean number of I/Os ($O_2$)*

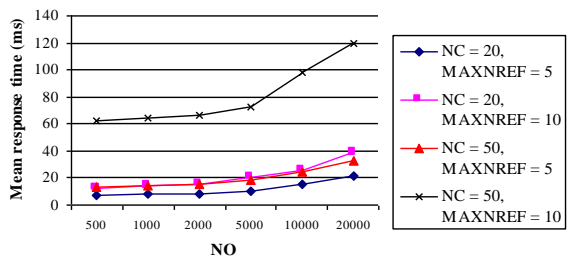
*Figure 5: Mean response time ($O_2$)*

We can see that the performances of $O_2$ logically decreased in the three following cases.

- *NC* increase — This was due to the structure of the OCB schema. The more classes it contained, the deeper the inheritance graph was. Since information is inherited at each level from the upper level, leaf classes in the inheritance graph have bigger instances than root classes. Hence, a higher number of classes induced bigger object sizes, so the database occupied more disk pages.
- *MAXNREF* increase — The number of objects accessed by transactions that browsed all the references increased.
- *NO* increase — The database got bigger and objects were distributed over more disk pages.

The evolution of our two performance criteria was quite similar. This result was expected, since most treatments performed by the system when running OCB deal with loading objects from disk.

Results for Texas

*Material Conditions*

Texas version 0.5 was installed on a PC Pentium-II 266 with a 64 MB SDRAM. The host operating system was Linux 2.0.30. The swap partition size was 64 MB. Texas has been compiled with the GNU C++ compiler version 2.7.2.1.

*Object Base Generation*

Figure 6 displays the average time for database generation function of the number of instances and the number of classes in the database. It shows that generation time did not increase linearly. However, the longest generation times were approximately 10 minutes long, which was an acceptable rate.

Texas did not appear to have the same behavior than $O_2$ because average generation time was greater when the schema contained few classes. This result can be attributed to two phenomena.

- The graph consistency check for acyclic graphs was more complex when the number of classes was low. In these conditions, the interclass references were dispersed in a reduced class interval and formed very dense graphs.
- When the database did not fit wholly into the main memory, the system swapped, which was costly both in terms of I/Os and time.

The actual size of the object bases generated with Texas was always less than 60 MB, as shown in Figure 8, allowing them to be stored in the 64 MB memory. Hence, the graph consistency check was prevalent while in the case of $O_2$, swap was prevalent. This hypothesis has been checked with Texas by reducing the available memory under Linux to 16 MB. Figure 7 displays the results of these tests, which confirmed our assumption.

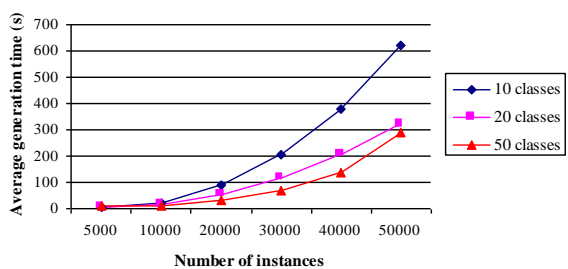
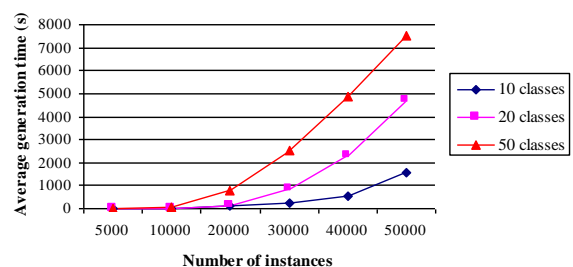

*Figure 6: Database generation time (Texas)*      *Figure 7: DB generation time with 16 MB memory*

Figure 8 eventually shows how the database real size evolved with the number of instances and the number of classes in the database. As happened with $O_2$, this evolution was linear. The average database size was about 20 MB with Texas. The object bases generated with $O_2$ were one third bigger due to the objects storage format: Texas directly uses the memory format while $O_2$ uses the WiSS (Chou, 1985) record structures that are more elaborate.

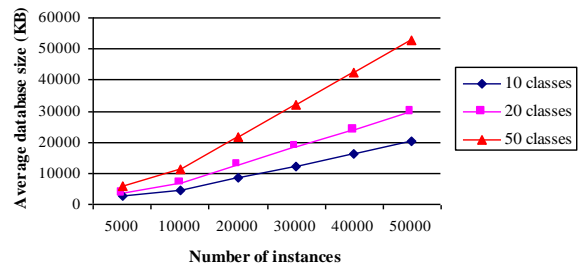

*Figure 8: Actual database size (Texas)*

*Object Base Usage*

In Figure 9, we plotted the mean number of I/Os globally necessary to execute the transactions function of the number of instances in the object base (*NO*), for our four database configurations. We did the same in Figure 10 for the mean response time.

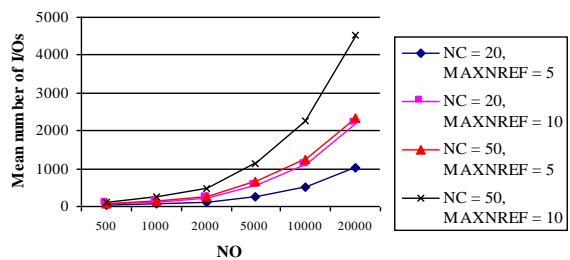 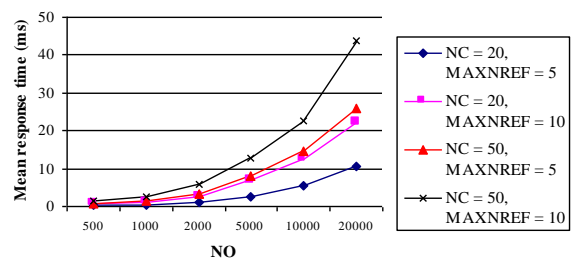

*Figure 9: Mean number of I/Os (Texas)*  *Figure 10: Mean response time (Texas)*

In the case of Texas, the correlation between the mean number of I/Os and the mean response time appeared tighter than for $O_2$. $O_2$ indeed includes many more features than Texas (security, concurrency control, and others) that add an overhead that is not directly linked to disk accesses.

Results for DSTC/Texas

The transactions selected for this series of experiments were depth-3 hierarchy traversals and depth-2 simple traversals. The depth of traversals was reduced regarding OCB's default parameters so that the generated clusters were not too big and the effects of clustering were clear. Technical problems were also encountered when the database size increased and DSTC attempted to build too large clusters. The traversals have been performed from 100 predefined root objects and each of them was executed 10 times.

Table 6 displays mean numbers of I/Os and response times concerning database usage before and after clustering. Our results showed that for both the transaction types used, the DSTC clustering technique allowed substantial increases in performances. The gain factor was about 5 for a medium object base and about 30 for a large one. We had the confirmation that the effects of clustering were stronger when the database size was greater than the memory size. Indeed, the smaller the database size, the more the system has to perform page replacements. Unused pages do not normally remain in memory for long.

Clustering overhead does not appear in the "large" base column because the medium base was reused (both the initial and the clustered configurations) with a reduced amount of memory. The results obtained show that this overhead was very important both in terms of time and I/Os. This is actually why techniques such as DSTC are usually triggered when the database is idle. Furthermore, reclustering the database is not a usual operation: a given object clustering may be employed during several sessions before being reconsidered. It is then important to determine the period after which clustering becomes advantageous, i.e., the time after which the induced overhead becomes lower than the achieved performance increase.

|  | **Medium base** | | | | **"Large" base** | | | |
|---|---|---|---|---|---|---|---|---|
|  | **Hierarchy traversals** | | **Simple traversals** | | **Hierarchy traversals** | | **Simple traversals** | |
| Pre-clustering usage | 1890.7 | 17.7 | 1837.4 | 15.7 | 12504.6 | 102.1 | 12068.1 | 103.1 |
| Post-clustering usage | 330.6 | 3.3 | 313.1 | 3.0 | 424.3 | 2.9 | 401.3 | 2.7 |
| *Gain factor* | *5.7* | *5.4* | *5.9* | *5.2* | *29.5* | *35.2* | *30.1* | *38.7* |
| Clustering overhead | 12799.6 | 125.8 | 12708.8 | 124.3 | | | | |

*Table 6: Effect of DSTC on Texas' performances (mean number of I/Os / mean response time, in ms)*

CONCLUSIONS AND FUTURE RESEARCH

We have presented in this paper the full specifications for a new object-oriented benchmark: OCB. Its main qualities are its richness, its flexibility, and its compactness. OCB indeed offers an object base whose complexity has never been achieved before in object-oriented benchmarks. Furthermore, since this database and likewise the transactions running on it are wholly tunable through a collection of comprehensive but easily set parameters, OCB can be used to model many kinds of object-oriented database applications. Eventually, OCB's code is short, reasonably easy to implement, and easily portable.

We have shown our benchmark was merely feasible by measuring generation time for its random database. It appears that in the worst case, an OCB object base is generated in less than one hour, which is quite acceptable. Furthermore, the largest databases can be saved for multiple uses.

We have also illustrated the genericity of our benchmark by showing how it could imitate both the schema and the operations of four existing benchmarks. The flaws identified in these previous benchmarks have been underlined and an attempt was made to correct them. We eventually demonstrated that OCB could be used as a general-purpose benchmark by evaluating the performances of the $O_2$ OODB and the Texas persistent object store. We also showed it could serve as a more specialized benchmark by testing the effects of the DSTC clustering method on the performances of Texas.

Future work concerning this study chiefly concerns the actual exploitation of OCB. We plan to benchmark several different systems featuring clustering techniques or not, for the sake of performance comparison or to determine if their configuration fits a certain purpose. Other aspects of OODB performance could also be tested, like buffering or indexing.

Future research about the OCB benchmark itself is mainly divided into two axes. First, we only exposed the principles of a multi-user version of our benchmark. The transition from the single-user version toward an operational multi-user version is not immediate and requires a particular care. The aim of this evolution is to evaluate the efficiency of concurrency control and to see how systems react when faced to a more important and heterogeneous workload. Since OODBs normally operate in a concurrent environment, their performances cannot be gauged with a single-user benchmark.

Second, one very different aspect we did not consider yet is the "qualitative" element that is important to take into account when selecting an OODB. Atkinson, Birnie, Jackson, and Phil-brow (1992), Banerjee and Gardner (1995), Kempe, Kowarschick, Kießling, Hitzelgerger, and Dutkowski (1995) all insist on the fact that functionality is at least as important as raw performances. Hence, criteria concerning these functionalities should be worked out. Sheer performance could be viewed as one of these criteria. Concerning optimization methods, we could, for instance, evaluate if a clustering heuristic's parameters are easy to apprehend and set up or if the algorithm is easy to use or transparent to the user.

Eventually, another point that can be considered is the adequacy of OCB to evaluate the performances of object-relational systems. Our generic model can of course be implemented with an object-relational system and most the operations are relevant for such a system. Thus, OCB can allow the comparison of different logical or physical organizations (distribution of the objects into tables, implementation of associations by values or by pointers, distribution of tables into tablespaces, index…). OCB can be considered as a candidate benchmark for this type of systems, even if extensions are needed to take into account additional aspects, regarding Abstract Data Types, in particular.

ACKNOWLEDGEMENTS

The authors would like to thank the editor and anonymous referees for their thoughtful criticisms and suggestions, through which this paper was greatly improved.